\documentclass{article}
\usepackage[dblblindworkshop, final]{neurips_2026}
\workshoptitle{Machine Learning and the Physical Sciences}

\usepackage[utf8]{inputenc}
\usepackage[T1]{fontenc}
\usepackage{hyperref}
\usepackage{float}
\usepackage{url}
\usepackage{booktabs}
\usepackage{amsfonts}
\usepackage{nicefrac}
\usepackage{microtype}
\usepackage{xcolor}
\usepackage{graphicx}
\newcommand{\arcsec}{\ensuremath{''}}

\makeatletter
\renewcommand{\@noticestring}{Accepted to the 9th Workshop on Machine Learning and the Physical Sciences (ML4PS 2026).}
\makeatother

\title{Identifying Scientifically Interesting Galaxies with Vision-Language Models}

\author{%
  Yixuan Cheng \\
  University of Toronto
  \And
  Nolan Koblischke \\
  University of Toronto
  \And
  Jo Bovy \\
  University of Toronto
}

\begin{document}

\maketitle

\begin{abstract}
Anomaly searches in large astronomical image datasets aim to identify rare objects for follow-up investigation. Yet the objects that are most statistically unusual are not necessarily those of greatest scientific relevance. We investigate whether pretrained vision-language models can directly assess scientific interest from natural-language instructions, enabling galaxy images to be ranked without labeled examples. We evaluate models on a 20,000-image Hubble Space Telescope sample that includes 167 reference anomalies previously identified through machine learning and expert inspection. By scoring images for scientific interest, Gemini 3.1 Flash-Lite places 98.2\% of the reference anomalies within the highest-ranked $\sim$2,000 images, at a present-day inference cost of approximately \$5.05 per 10,000 images. We recover known gravitational lenses, jellyfish-galaxy candidates, and galaxies with visible jets, and release 102 high-scoring candidates for further investigation. These results show that an off-the-shelf vision-language model, without any training on the target data, offers a practical way to rank large image datasets by scientific relevance rather than rarity. The project code is available at \url{https://github.com/TheCheng2005/Galaxy-Anomaly-Search/}.
\end{abstract}

\section{Introduction}

Modern astronomical imaging surveys make data volume a limiting factor in discovery. New surveys such as Euclid~\citep{euclidredbook} will image over a billion galaxies. Deciding which images warrant astronomers’ attention is a triage problem, often approached through anomaly detection, which identifies unusual objects with scientifically interesting features that may not be known a priori.

The classic approach to anomaly detection is to use learned representations and assign an anomaly score to each object by its distance from the bulk of the data. For example, \citet{storeyfisher2021} trained a generative adversarial network on nearly a million galaxy images and identified anomalies from reconstruction errors and unusual learned features. \citet{lochner2021} used active learning to incorporate human feedback into anomaly rankings, while \citet{mantha2024} combined machine-learned representations with volunteer assessments of galaxy images. Most recently, AnomalyMatch \citep{gomez2025} combines a semi-supervised algorithm with active learning. \citet{oryan2025} applied AnomalyMatch to identify anomalies in 99.6 million Hubble Space Telescope (HST) cutouts.

However, the image-based anomaly detection methods described above derive their rankings primarily from visual features and therefore lack broader scientific context. A distance-based algorithm might fail to distinguish between rare scientific phenomena and rare instrumental artifacts, since both can lie far from the bulk of the data in the model's representation space~\citep{lochner2021,etsebeth2024}. The supervised methods get around this by training on human-labeled examples of anomalies, but this misses unknown-unknowns and undiscovered rare objects theorized in the literature.

Vision-language models (VLMs; e.g. Gemini~\citep{gemini} and Qwen~\citep{bai2025qwen3vltechnicalreport}) offer a way to bring scientific context into image-based anomaly detection. These models combine a language model with a vision encoder and are trained on large collections of image captions and questions, enabling image interpretation alongside natural-language instructions. Exposure to scientific text and images~\citep{mint} could help these models distinguish instrumental artifacts from scientifically interesting objects.

VLMs have recently been shown to describe and classify galaxy images with moderate success~\citep{Koblischke_2026,chiaberge2026}. Their ability to find astronomical anomalies has also been explored by \citet{pesta2026}, who used VLM judgments of stellar brightness variations to guide an active-learning search for unusual sources, and by \citet{surakan2026evaluatingzeroshotanomalydetection}, who applied a zero-shot anomaly detector built on a contrastive vision-language model to James Webb Space Telescope galaxy images.

In this work, we investigate whether Gemini~\citep{gemini} and Qwen~\citep{qwen3.5} VLMs can identify scientifically interesting anomalies in galaxy images without examples or task-specific fine-tuning, using their scientific judgments directly to rank galaxy images for human inspection. We compare strategies on 20,000 Hubble Space Telescope images, measuring recovery of reference anomalies previously identified by AnomalyMatch~\citep{gomez2025} and expert inspection. We then examine high-scoring sources outside the reference catalog, recovering previously studied objects and assembling a catalog of candidates for further investigation.

\section{Method}
\label{sec:method}

\paragraph{Dataset Construction} We evaluate anomaly recovery using the reference sample of \citet{oryan2025}, who applied AnomalyMatch~\citep{gomez2025}, a semi-supervised active-learning method, to 99.6 million Hubble Space Telescope (HST) images. They identified 1,176 anomalies after expert inspection and de-duplication of the 5,000 highest-scoring cutouts. We rebuild a comparable pool from the same HST F814W ACS/WFC dataset: we sample 10 million of the 77,651,429 extended-source detections in the Hubble Source Catalog v3~\citep{Whitmore2016}, then keep cutouts separated by at least $10\arcsec$, preferring unsaturated sources without quality flags. The surviving 223,195 detections each become a $150\times150$\,px cutout, $7.5\arcsec$ per side, scaled as in AnomalyMatch to preserve visible morphological features across intensity ranges. Cutouts receive an \emph{anomaly} label when an AnomalyMatch detection lies within $3\arcsec$, leading to 167 anomalies in our benchmark set. To limit inference costs, we use a 20,000-image subset containing all reference anomalies and randomly select 19,833 unlabeled background cutouts from 223,195 surviving detections. We publicly release this benchmark dataset.\footnote{\url{https://huggingface.co/datasets/astronolan/hubble-anomaly}}

\paragraph{Image Scoring} We compare two prompting strategies for ranking images by scientific interest: \textit{Likert} and \textit{Tournament}. In \textit{Likert}, the model evaluates a $4\times4$ grid, selects interesting images, and assigns each selected image a score from 1 to 5, while unselected images receive 0. Each image is evaluated in ten randomly assembled grids, and its scores are summed for a final score from 0 to 50. In \textit{Tournament}, the model evaluates a $2\times2$ grid and retains any number of images for the next round. Retained images are randomly regrouped, while discarded images are excluded from subsequent rounds. After ten rounds, each image receives a score from 0 to 10 corresponding to the round it reached. In both methods, each grid is submitted in a separate API request containing no information from previous rounds. The exact prompts used for these methods are provided in Appendix~\ref{app:prompts}.

We assess what fraction of the reference anomalies each method recovers. Because \textit{Tournament} retains approximately 2,000 images after ten rounds, we choose \textit{Likert} thresholds that select similarly sized sets. We report recall for these complete threshold-selected sets and expected Recall@2000, assuming random ordering within tied scores, for comparison at an identical inspection budget.

\paragraph{Vision-Language Models} We select Gemini 3.1 Flash-Lite for its low inference cost and strong reported performance on visual-understanding benchmarks~\citep{gemini31flashlite2026}. Qwen3.5-397B\footnote{Accessed via \url{https://openrouter.ai/}} provides an open-weight comparison from a widely studied model family~\citep{yin2024survey}, with previous applications to astronomical image analysis~\citep{Koblischke_2026, drozdova2025radio}. Both models support reasoning, which we disable for Qwen and set to low for Gemini to reduce inference costs, motivated by reports of limited or inconsistent benefits on visual tasks~\citep{2025arXiv251103328H}. For Gemini \textit{Likert}, increasing thinking level from low to high changed recovery negligibly (164 vs. 163 anomalies) but with a higher cost (\$5.05 vs. \$17.23 per 10,000 images).

\paragraph{Conventional Baselines} For comparison, we embed the same 20,000 cutouts with Zoobot~\citep{zoobot}, a family of galaxy foundation models pretrained to predict galaxy morphologies \citep{walmsley2024scaling}. Its training set, GZ Evo~\citep{walmsley2025gzevo}, holds over 100 million volunteer labels for 823,000 galaxy images from five Galaxy Zoo campaigns on four telescopes, including Galaxy Zoo: Hubble, which classified galaxies in HST legacy imaging~\citep{willett2017gzh}. We use the 15.6M-parameter ConvNeXt-nano encoder in its single-channel version, which matches our single-band cutouts\footnote{\url{https://huggingface.co/mwalmsley/zoobot-encoder-greyscale-convnext_nano}}. Each cutout is upsampled to $224\times224$\,px and represented by the encoder's 640 pooled features. We obtain anomaly scores on the images by two unsupervised methods: the mean cosine distance to their 10 nearest neighbours (kNN distance), and an isolation forest~\citep{liu2008isolation} of 1,000 trees, each grown on 256 randomly sampled images. We evaluate the 2,000 highest-scoring images from each. These are common defaults, but we also experimented with more neighbours, more trees, and different numbers of sampled images. None of the new parameters changed the results by more than 2 percentage points.

\section{Results}
\label{sec:results}

\begin{figure} [H]
    \centering
    \includegraphics[width=0.96\linewidth]{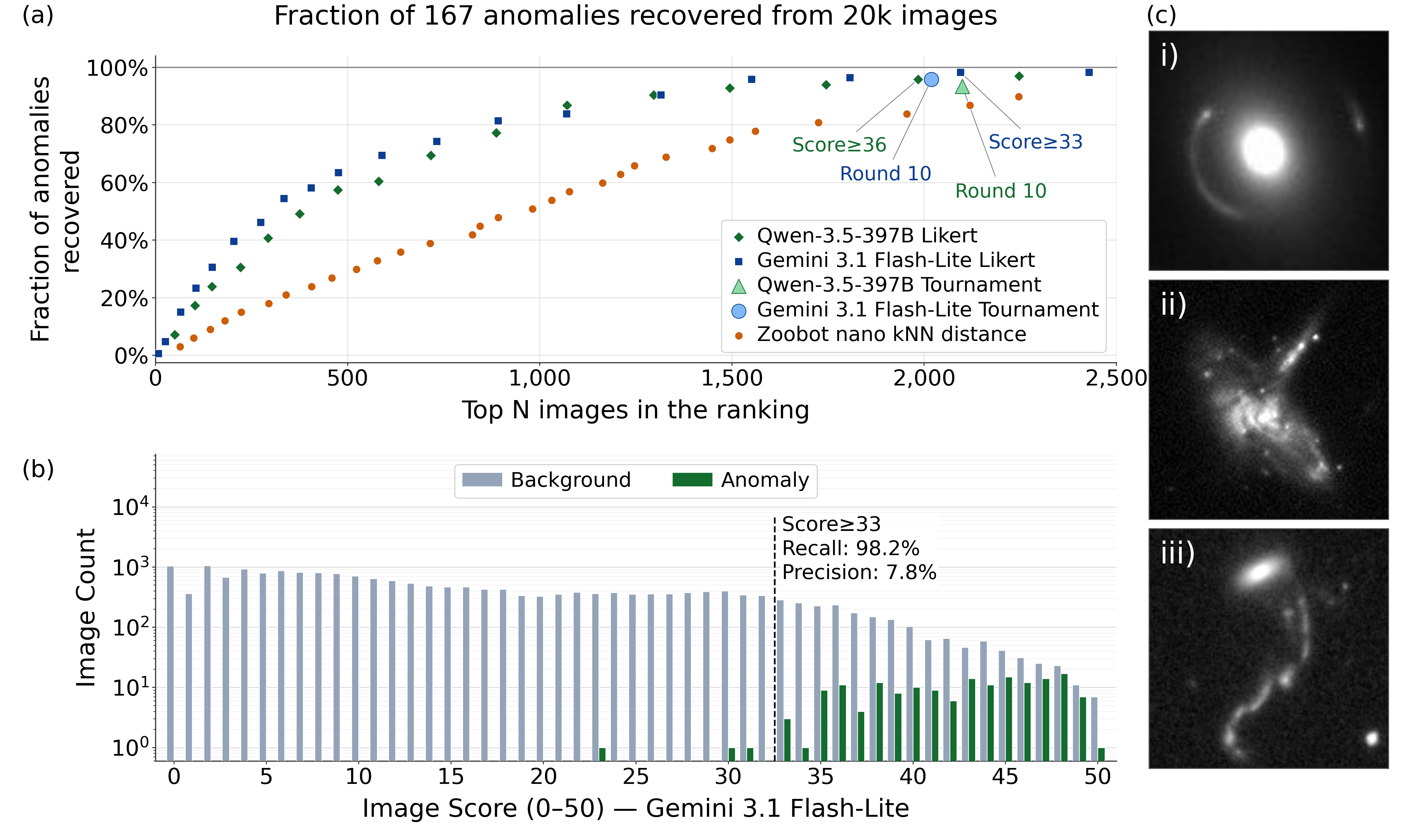}
    \caption{(a) Fraction of the reference anomalies recovered versus the number of images selected by a score threshold. Each \textit{Likert} point includes all images at or above a given score threshold. Thresholds $\geq33$ for Gemini and $\geq36$ for Qwen are shown. Each \textit{Tournament} method is displayed for the final Round 10. We also plot Zoobot nano embeddings ranked by kNN distance, which show a more linear trend.  (b) Gemini 3.1 Flash-Lite \textit{Likert} score distributions for 20,000 images. The dashed boundary marks score $\geq33$, selecting 2,095 images with recall 98.2\% and precision 7.8\%. (c) Three high-scoring Gemini \textit{Likert} images: i) the strong gravitational lens SDSS~J1205+4910 \citep{Bolton2006} (RA=181.418362$^\circ$, Dec=49.174771$^\circ$, 50/50), ii) a disturbed galaxy (RA=202.009029$^\circ$, Dec=$-$31.515268$^\circ$, 49/50), and iii) a tidal feature or jellyfish galaxy (RA=334.403219$^\circ$, Dec=0.835206$^\circ$, 49/50). Only ii) is identified by AnomalyMatch.}
    \label{fig:3panels}
\end{figure}

\begin{table}[H]
  \centering
  \footnotesize
  \setlength{\tabcolsep}{4pt}
\begin{tabular*}{\textwidth}{@{\extracolsep{\fill}}llcrcccr@{}}
    \toprule
    Model & Method & Threshold & N Selected & Recall & Precision & Recall@2000 & \$/10,000 \\
    & & & & & & (accounting for ties) & Images \\
    \midrule
    Gemini 3.1 Flash-Lite & \textit{Likert} & 33/50 & 2{,}095 & 0.982 & 0.078 & 0.976 & \$5.05 \\
    Qwen3.5-397B-A17B & \textit{Likert} & 36/50 & 1{,}984 & 0.958 & 0.081 & 0.959 & \$6.69 \\
    Gemini 3.1 Flash-Lite & \textit{Tournament} & Round 10 & 2{,}019 & 0.958 & 0.079 & 0.949 & \$3.66 \\
    Qwen3.5-397B-A17B & \textit{Tournament} & Round 10 & 2{,}099 & 0.934 & 0.074 & 0.890 & \$1.80 \\
    \midrule
    Zoobot ConvNeXt-nano & \textit{kNN distance} & Top 2{,}000 & 2{,}000 & 0.850 & 0.071 & 0.850 & -- \\
    Zoobot ConvNeXt-nano & \textit{Isolation forest} & Top 2{,}000 & 2{,}000 & 0.701 & 0.059 & 0.701 & -- \\
    \bottomrule
\end{tabular*}
\caption{Performance in recovering the 167 AnomalyMatch reference anomalies from the 20,000-image sample. N Selected is the number of images at or above the specified threshold. Recall and precision are measured on these selected sets. Expected Recall@2000 gives an equal set size comparison assuming random ordering within tied scores. Costs are USD per 10,000 images.}
  \label{tab:recall}
\end{table}

\paragraph{Anomaly Recovery}

Figure~\ref{fig:3panels}(a) shows that vision-language models can accurately recover interesting anomalies, with Gemini \textit{Likert} recovering 49.7\% of the reference anomalies in its top-300 ranked images and nearly all in its top 2,000. Gemini \textit{Likert} achieves slightly higher recall than \textit{Tournament}, recovering 98.2\% of the reference anomalies among 2,095 selected images, compared with 95.8\% among 2,019 images. Qwen \textit{Likert} performs similarly to Gemini, recovering 95.8\% of the reference anomalies among 1,984 selected images, while Qwen \textit{Tournament} recovers 93.4\% among 2,099 images. The conventional baselines recover fewer among their top 2,000 images. Zoobot kNN distance finds 85.0\% of the reference anomalies and the isolation forest finds 70.1\%. The gap is larger for the top 300 images, where they recover only 18.0\% and 10.8\% respectively.

Table~\ref{tab:recall} shows \textit{Tournament} is substantially cheaper than \textit{Likert} at similar recall: \$3.66 vs. \$5.05 per 10,000 images for Gemini (27\% less) and \$1.80 vs. \$6.69 for Qwen (73\% less), in both cases recovering slightly fewer reference anomalies. The savings come from \textit{Tournament} eliminating most images after round one (Gemini discards 11,974 of 20,000; Qwen discards 13,178), leaving far fewer images to score in later rounds, whereas \textit{Likert} rescores every image 10 times. \textit{Likert} yields 51 distinct scores (Fig.~\ref{fig:3panels}b), enabling fine control over the threshold and therefore the precision/recall trade-off. \textit{Tournament} scores are much coarser, relaxing the threshold for Gemini from Round 10 to Round 9 adds 118 images while recovering only one additional reference anomaly. Notably, all 167 survive the first \textit{Tournament} round and none receives a final \textit{Likert} score of zero (Fig.~\ref{fig:3panels}b).

To assess sensitivity to background sampling and grid randomization, we repeat Gemini \textit{Likert} with two additional seeds. Across all three runs, recall at the same score threshold $\geq33$ is $0.980$ on average with a standard deviation of $\pm0.015$, selecting 2,082--2,148 images per run. We also test whether the models avoid ranking instrumental artifacts as scientifically interesting. Manual inspection of the 200 highest-ranked Gemini \textit{Likert} images finds only four images whose dominant feature we determine to be an instrumental artifact.

\paragraph{High-Scoring Sources beyond the Reference}

Since the visual inspection by \citet{oryan2025} was restricted to the 5,000 highest-ranked images, the absence of an AnomalyMatch label does not exclude a source from being scientifically interesting. We search for positional counterparts within $3\arcsec$ in a dataset of galaxy mentions in astronomical papers~\citep{koblischke2026galaxywiki},\footnote{\url{https://huggingface.co/datasets/astronolan/galaxy-mentions}} and the interacting-galaxy catalog of \citet{ORyan_2023}. Of the 138 images with Gemini \textit{Likert} scores $\geq45$ outside the AnomalyMatch reference catalog, 33 match the interacting-galaxy catalog and one additional image matches the galaxy-mentions dataset.

Our screening recovers several previously studied objects, including the strong gravitational lens SDSS~J1205+4910 \citep{Bolton2006} notably missing from the AnomalyMatch catalog (Fig.~\ref{fig:3panels}(c)i, score 50/50), Arp~238-E, the eastern member of a strongly interacting spiral pair in the Atlas of Peculiar Galaxies (score 50/50) \citep{Arp_1966, Xu_2021}, and IC~5298, classified as a radio AGN (score 45/50) \citep{Vardoulaki_2015}. The source shown in Fig.~\ref{fig:3panels}(c)iii (score 49/50) has asymmetric features suggestive of tidal disruption and matches an entry in
an interacting galaxy catalog~\citep{ORyan_2023}. The top-200 images ranked by Gemini \textit{Likert} also include sources classified in the AnomalyMatch reference catalog as jet candidates, mergers, gravitational lenses, jellyfish galaxies, and ring galaxies.

After these catalog exclusions, 104 sources remain. We exclude two additional sources identified as previously studied through a manual review of the literature, leaving 102 candidates. We release this catalog at \url{https://github.com/TheCheng2005/Galaxy-Anomaly-Search/blob/main/paper_results/catalogue/candidates.csv} and the 30 highest-scoring candidates are shown in Appendix~\ref{app:candidates}.

\section{Discussion \& Conclusion}
\label{sec:discussion}

Our results show that VLMs can sift through large astronomical image datasets to uncover scientifically interesting sources that merit further study, providing a scalable route for prioritizing follow-up in largely unexplored image archives. The rediscovery of known strong lenses, interacting galaxies, and a radio AGN demonstrates recovery beyond the reference sample, while our candidate catalog of 102 high-scoring sources provides targets for further investigation (Appendix~\ref{app:candidates}). This capability extends to open-weight models, as Qwen can run on academic GPU clusters and nearly matches the closed-source Gemini model. At a cost of \$180 per million images, the Qwen \textit{Tournament} method could feasibly be used to explore the datasets produced by wide-field telescope imaging surveys.

We note two important limitations of our evaluation. First, for evaluation cost considerations, the benchmark retains all available reference anomalies while subsampling the background, so the reported precision is specific to this enriched sample and may differ in a blind search. To estimate performance at a more realistic ratio of one anomaly per 1,200 images, we rescale the score distribution of Fig.~\ref{fig:3panels}b instead of scoring more images. The background count in each score bin is multiplied by ten, and the anomaly counts are left unchanged. Recall at the threshold $\geq$ 33 stays at 98.2\%, because the additional background images add to neither the true positives nor the false negatives counts. However, now 19,474 images instead of 2,095 are above the threshold, and precision falls from 7.8\% to 0.84\%, meaning that experts must examine ten times more images to find the same number of anomalies. For the top-ranked 2,000 images, expected recall decreases from 97.6\% to 45.6\%. Second, the reference labels are incomplete, meaning that some high-scoring sources counted as background may in fact be scientifically interesting.

Future work will investigate cost-efficient local deployment, astronomy-specific fine-tuning, and evaluation on more diverse datasets. We also plan to apply this methodology to an upcoming large-scale survey release such as Euclid DR1~\citep{euclidredbook}, where it can be tested on previously unseen survey data for its ability to surface novel phenomena.

\begin{ack}
This work used API credits provided by Google through its Gemini Academic Program. Author NK would like to acknowledge the support of the Natural Sciences and Engineering Research Council of Canada (NSERC) [funding reference number 721697856] and the Data Sciences Institute at the University of Toronto through grant number DSI-DSFY4R1P01. Author YC would like to acknowledge the support of the Data Sciences Institute at the University of Toronto for the SUDS program.

This work makes use of observations made with the NASA/ESA Hubble Space Telescope, and obtained from the Hubble Legacy Archive, which is a collaboration between the Space Telescope Science Institute (STScI/NASA), the Space Telescope European Coordinating Facility (ST-ECF/ESAC/ESA) and the Canadian Astronomy Data Centre (CADC/NRC/CSA). This research has also made use of the NASA/IPAC Extragalactic Database, which is funded by the National Aeronautics and Space Administration and operated by the California Institute of Technology.

\end{ack}

\bibliographystyle{plainnat}
\bibliography{references}

@ARTICLE{storeyfisher2021,
       author = {{Storey-Fisher}, Kate and {Huertas-Company}, Marc and {Ramachandra}, Nesar and {Lanusse}, Francois and {Leauthaud}, Alexie and {Luo}, Yifei and {Huang}, Song and {Prochaska}, J. Xavier},
        title = "{Anomaly detection in Hyper Suprime-Cam galaxy images with generative adversarial networks}",
      journal = {Monthly Notices of the Royal Astronomical Society},
         year = 2021,
        month = dec,
       volume = {508},
       number = {2},
        pages = {2946-2963},
          doi = {10.1093/mnras/stab2589},
archivePrefix = {arXiv},
       eprint = {2105.02434},
 primaryClass = {astro-ph.GA},
       adsurl = {https://ui.adsabs.harvard.edu/abs/2021MNRAS.508.2946S}
}

@ARTICLE{lochner2021,
       author = {{Lochner}, M. and {Bassett}, B.~A.},
        title = "{ASTRONOMALY: Personalised active anomaly detection in astronomical data}",
      journal = {Astronomy and Computing},
         year = 2021,
        month = jul,
       volume = {36},
          eid = {100481},
        pages = {100481},
          doi = {10.1016/j.ascom.2021.100481},
archivePrefix = {arXiv},
       eprint = {2010.11202},
 primaryClass = {astro-ph.IM},
       adsurl = {https://ui.adsabs.harvard.edu/abs/2021A&C....3600481L}
}

@ARTICLE{etsebeth2024,
       author = {{Etsebeth}, V. and {Lochner}, M. and {Walmsley}, M. and {Grespan}, M.},
        title = "{Astronomaly at scale: searching for anomalies amongst 4 million galaxies}",
      journal = {Monthly Notices of the Royal Astronomical Society},
         year = 2024,
        month = mar,
       volume = {529},
       number = {1},
        pages = {732-747},
          doi = {10.1093/mnras/stae496},
archivePrefix = {arXiv},
       eprint = {2309.08660},
 primaryClass = {astro-ph.IM},
       adsurl = {https://ui.adsabs.harvard.edu/abs/2024MNRAS.529..732E}
}

@ARTICLE{oryan2025,
       author = {{O'Ryan}, David and {G{\'o}mez}, Pablo},
        title = "{Identifying astrophysical anomalies in 99.6 million source cutouts from the Hubble legacy archive using AnomalyMatch}",
      journal = {Astronomy and Astrophysics},
         year = 2025,
        month = dec,
       volume = {704},
          eid = {A227},
        pages = {A227},
          doi = {10.1051/0004-6361/202555512},
archivePrefix = {arXiv},
       eprint = {2505.03508},
 primaryClass = {astro-ph.IM},
       adsurl = {https://ui.adsabs.harvard.edu/abs/2025A&A...704A.227O}
}

@article{gomez2025,
    author = {Gómez, Pablo and Ruhberg, Laslo E and Nardone, Maria Teresa and O’Ryan, David},
    title = {AnomalyMatch: discovering rare objects of interest with semisupervised and active learning},
    journal = {RAS Techniques and Instruments},
    volume = {5},
    pages = {rzag045},
    year = {2026},
    month = {01},
    issn = {2752-8200},
    doi = {10.1093/rasti/rzag045},
    url = {https://doi.org/10.1093/rasti/rzag045},
    eprint = {https://academic.oup.com/rasti/article-pdf/doi/10.1093/rasti/rzag045/68608256/rzag045.pdf},
}

@ARTICLE{Whitmore2016,
       author = {{Whitmore}, Bradley C. and {Allam}, Sahar S. and {Budav{\'a}ri}, Tam{\'a}s and {Casertano}, Stefano and {Downes}, Ronald A. and {Donaldson}, Thomas and {Fall}, S. Michael and {Lubow}, Stephen H. and {Quick}, Lee and {Strolger}, Louis-Gregory and {Wallace}, Geoff and {White}, Richard L.},
        title = "{Version 1 of the Hubble Source Catalog}",
      journal = {The Astronomical Journal},
         year = 2016,
        month = jun,
       volume = {151},
       number = {6},
          eid = {134},
        pages = {134},
          doi = {10.3847/0004-6256/151/6/134},
archivePrefix = {arXiv},
       eprint = {1602.04861},
 primaryClass = {astro-ph.IM},
       adsurl = {https://ui.adsabs.harvard.edu/abs/2016AJ....151..134W}
}

@article{mantha2024,
  author  = {Mantha, Kameswara Bharadwaj and Roberts, Hayley
             and Fortson, Lucy and Lintott, Chris
             and Dickinson, Hugh and others},
  title   = {Through the Citizen Scientists' Eyes: Insights into Using
             Citizen Science with Machine Learning for Effective
             Identification of Unknown-Unknowns in Big Data},
  journal = {Citizen Science: Theory and Practice},
  volume  = {9},
  number  = {1},
  pages   = {40},
  year    = {2024},
  doi     = {10.5334/cstp.740},
  url     = {https://doi.org/10.5334/cstp.740}
}

@ARTICLE{gemini,
       author = {{Comanici}, Gheorghe and {Bieber}, Eric and {Schaekermann}, Mike and others},
        title = "{Gemini 2.5: Pushing the Frontier with Advanced Reasoning, Multimodality, Long Context, and Next Generation Agentic Capabilities}",
      journal = {arXiv e-prints},
         year = 2025,
        month = jul,
          eid = {arXiv:2507.06261},
        pages = {arXiv:2507.06261},
          doi = {10.48550/arXiv.2507.06261},
archivePrefix = {arXiv},
       eprint = {2507.06261},
 primaryClass = {cs.CL},
       adsurl = {https://ui.adsabs.harvard.edu/abs/2025arXiv250706261C}
}

@misc{mint,
      title={MINT-1T: Scaling Open-Source Multimodal Data by 10x: A Multimodal Dataset with One Trillion Tokens}, 
      author={Anas Awadalla and Le Xue and Oscar Lo and Manli Shu and Hannah Lee and Etash Kumar Guha and Matt Jordan and Sheng Shen and Mohamed Awadalla and Silvio Savarese and Caiming Xiong and Ran Xu and Yejin Choi and Ludwig Schmidt},
      year={2024},
      eprint={2406.11271},
      archivePrefix={arXiv},
      primaryClass={cs.CV},
      url={https://arxiv.org/abs/2406.11271}, 
}

@ARTICLE{2025arXiv251103328H,
       author = {{Hong}, Jindong and {Chen}, Tianjie and {Luo}, Lingjie and {Zheng}, Chuanyang and {Xu}, Ting and {Yu}, Haibao and {Qiu}, Jianing and {Chen}, Qianzhong and {Huang}, Suning and {Xu}, Yan and {Gui}, Yong and {He}, Yijun and {Sun}, Jiankai},
        title = "{Benchmarking the Thinking Mode of Multimodal Large Language Models in Clinical Tasks}",
      journal = {arXiv e-prints},
         year = 2025,
        month = nov,
          eid = {arXiv:2511.03328},
        pages = {arXiv:2511.03328},
          doi = {10.48550/arXiv.2511.03328},
archivePrefix = {arXiv},
       eprint = {2511.03328},
 primaryClass = {cs.CL},
       adsurl = {https://ui.adsabs.harvard.edu/abs/2025arXiv251103328H}
}

@misc{bai2025qwen3vltechnicalreport,
      title={Qwen3-VL Technical Report}, 
      author={Shuai Bai and Yuxuan Cai and Ruizhe Chen and Keqin Chen and others},
      year={2025},
      eprint={2511.21631},
      archivePrefix={arXiv},
      primaryClass={cs.CV},
      url={https://arxiv.org/abs/2511.21631}, 
}

@misc{chiaberge2026,
      title={Vision-Language Model Ensembles Achieve Human-Expert Accuracy for Galaxy Merger Classification}, 
      author={Marco Chiaberge and Elias Stengel-Eskin and Massimo Stiavelli and Colin Norman},
      year={2026},
      eprint={2606.00415},
      archivePrefix={arXiv},
      primaryClass={astro-ph.GA},
      url={https://arxiv.org/abs/2606.00415}, 
}

@ARTICLE{euclidredbook,
       author = {{Laureijs}, R. and {Amiaux}, J. and {Arduini}, S. and {Augu{\`e}res}, J. -L.  and others},
        title = "{Euclid Definition Study Report}",
      journal = {arXiv e-prints},
         year = 2011,
        month = oct,
          eid = {arXiv:1110.3193},
        pages = {arXiv:1110.3193},
          doi = {10.48550/arXiv.1110.3193},
archivePrefix = {arXiv},
       eprint = {1110.3193},
 primaryClass = {astro-ph.CO},
       adsurl = {https://ui.adsabs.harvard.edu/abs/2011arXiv1110.3193L}
}

@misc{pesta2026,
      title={Agentic Active Learning Meets Visual Embeddings: Finding Anomalies among 370 000 Variable Stars from ASAS-SN}, 
      author={Milan Pesta and Yuan-Sen Ting},
      year={2026},
      eprint={2608.23688},
      archivePrefix={arXiv},
      primaryClass={astro-ph.SR},
      url={https://arxiv.org/abs/2608.23688}, 
}

@misc{surakan2026evaluatingzeroshotanomalydetection,
      title={Evaluating Zero-Shot Anomaly Detection for Identifying Unusual Galaxy Morphologies in CEERS and NGDEEP}, 
      author={T. B. Surakan and J. S. Kartaltepe and R. Pattnaik and C. K. Binu and I. G. Cox and T. Y. Hsiao and R. A. Lucas and F. Pacucci and G. Troiani and L. Y. Aaron Yung},
      year={2026},
      eprint={2610.29868},
      archivePrefix={arXiv},
      primaryClass={astro-ph.GA},
      url={https://arxiv.org/abs/2610.29868}, 
}

@article{Koblischke_2026,
              doi={10.3847/1538-4357/ae7b22},
              url={https://doi.org/10.3847/1538-4357/ae7b22},
              year={2026},
              month={jul},
              publisher={The American Astronomical Society},
              volume={1005},
              number={2},
              pages={170},
              author={Koblischke, Nolan and Parker, Liam and Lanusse, Francois and Bovy, Jo and Espejo, Irina and Ho, Shirley},
              title={Semantic Search for 100M+ Galaxy Images Using AI-generated Captions},
              journal={The Astrophysical Journal}
        }

@misc{gemini31flashlite2026,
  author       = {{Google DeepMind}},
  title        = {{Gemini 3.1 Flash-Lite Model Card}},
  year         = {2026},
  month        = mar,
  url          = {https://deepmind.google/models/model-cards/gemini-3-1-flash-lite/}
}

@article{yin2024survey,
  author       = {Yin, Shukang and Fu, Chaoyou and Zhao, Sirui
                  and Li, Ke and Sun, Xing and Xu, Tong
                  and Chen, Enhong},
  title        = {A survey on multimodal large language models},
  journal      = {National Science Review},
  year         = {2024},
  volume       = {11},
  number       = {12},
  pages        = {nwae403},
  doi          = {10.1093/nsr/nwae403}
}

@misc{drozdova2025radio,
  author       = {Drozdova, Mariia and Lastufka, Erica
                  and Kinakh, Vitaliy and Holotyak, Taras
                  and Schaerer, Daniel and Voloshynovskiy, Slava},
  title        = {Radio Astronomy in the Era of Vision-Language
                  Models: Prompt Sensitivity and Adaptation},
  year         = {2025},
  eprint       = {2509.02615},
  archivePrefix = {arXiv},
  primaryClass = {astro-ph.IM},
  doi          = {10.48550/arXiv.2509.02615},
  url          = {https://arxiv.org/abs/2509.02615}
}

@misc{qwen3.5,
  title  = {{Qwen3.5}: Towards Native Multimodal Agents},
  author = {{Qwen Team}},
  month  = {February},
  year   = {2026},
  url    = {https://qwen.ai/blog?id=qwen3.5}
}

@ARTICLE{Bolton2006,
       author = {{Bolton}, Adam S. and {Burles}, Scott and {Koopmans}, L{\'e}on V.~E. and {Treu}, Tommaso and {Moustakas}, Leonidas A.},
        title = "{The Sloan Lens ACS Survey. I. A Large Spectroscopically Selected Sample of Massive Early-Type Lens Galaxies}",
      journal = {The Astrophysical Journal},
         year = 2006,
        month = feb,
       volume = {638},
       number = {2},
        pages = {703-724},
          doi = {10.1086/498884},
archivePrefix = {arXiv},
       eprint = {astro-ph/0511453},
 primaryClass = {astro-ph},
       adsurl = {https://ui.adsabs.harvard.edu/abs/2006ApJ...638..703B}
}

@article{Xu_2021,
doi = {10.3847/1538-4357/ac0f77},
url = {https://doi.org/10.3847/1538-4357/ac0f77},
year = {2021},
month = {sep},
publisher = {The American Astronomical Society},
volume = {918},
number = {2},
pages = {55},
author = {Xu, C. K. and Lisenfeld, U. and Gao, Y. and Renaud, F.},
title = {NOEMA Observations of CO Emission in Arp 142 and Arp 238},
journal = {The Astrophysical Journal}
}

@ARTICLE{Arp_1966,
       author = {{Arp}, Halton},
        title = "{Atlas of Peculiar Galaxies}",
      journal = {The Astrophysical Journal Supplement Series},
         year = 1966,
        month = nov,
       volume = {14},
        pages = {1},
          doi = {10.1086/190147},
       adsurl = {https://ui.adsabs.harvard.edu/abs/1966ApJS...14....1A}
}

@article{ORyan_2023,
doi = {10.3847/1538-4357/acc0ff},
url = {https://doi.org/10.3847/1538-4357/acc0ff},
year = {2023},
month = {may},
publisher = {The American Astronomical Society},
volume = {948},
number = {1},
pages = {40},
author = {O’Ryan, David and Merín, Bruno and Simmons, Brooke D. and Vojteková, Antónia and Anku, Anna and Walmsley, Mike and Garland, Izzy L. and Géron, Tobias and Keel, William and Kruk, Sandor and Lintott, Chris J. and Mantha, Kameswara Bharadwaj and Masters, Karen L. and Reerink, Jan and Smethurst, Rebecca J. and Thorne, Matthew R.},
title = {Harnessing the Hubble Space Telescope Archives: A Catalog of 21,926 Interacting Galaxies},
journal = {The Astrophysical Journal}
}

@article{Vardoulaki_2015,
   title={Radio continuum properties of luminous infrared galaxies: Identifying the presence of an AGN in the radio},
   volume={574},
   ISSN={1432-0746},
   url={http://dx.doi.org/10.1051/0004-6361/201424125},
   DOI={10.1051/0004-6361/201424125},
   journal={Astronomy \& Astrophysics},
   publisher={EDP Sciences},
   author={Vardoulaki, E. and Charmandaris, V. and Murphy, E. J. and Diaz-Santos, T. and Armus, L. and Evans, A. S. and Mazzarella, J. M. and Privon, G. C. and Stierwalt, S. and Barcos-Muñoz, L.},
   year={2015},
   month=Jan, pages={A4} }

@article{zoobot,
  author  = {Walmsley, Mike and Allen, Campbell and Aussel, Ben and others},
  title   = {Zoobot: Adaptable Deep Learning Models for Galaxy Morphology},
  journal = {Journal of Open Source Software},
  year    = {2023},
  volume  = {8},
  number  = {85},
  pages   = {5312},
  doi     = {10.21105/joss.05312},
}

@inproceedings{liu2008isolation,
  author    = {Liu, Fei Tony and Ting, Kai Ming and Zhou, Zhi-Hua},
  title     = {Isolation Forest},
  booktitle = {2008 Eighth IEEE International Conference on Data Mining},
  year      = {2008},
  pages     = {413--422},
  doi       = {10.1109/ICDM.2008.17},
}

@misc{walmsley2024scaling,
  author        = {Walmsley, Mike and Bowles, Micah and Scaife, Anna M. M. and Makechemu, Jason Shingirai and Gordon, Alexander J. and Ferguson, Annette M. N. and Mann, Robert G. and Pearson, James and Popp, J{\"u}rgen J. and Bovy, Jo and Speagle, Josh and Dickinson, Hugh and Fortson, Lucy and G{\'e}ron, Tobias and Kruk, Sandor and Lintott, Chris J. and Mantha, Kameswara and Mohan, Devina and O'Ryan, David and Slijepcevic, Inigo V.},
  title         = {Scaling Laws for Galaxy Images},
  year          = {2024},
  eprint        = {2404.02973},
  archivePrefix = {arXiv},
  primaryClass  = {cs.CV},
  doi           = {10.48550/arXiv.2404.02973},
  url           = {https://arxiv.org/abs/2404.02973},
}

@article{willett2017gzh,
  author  = {Willett, Kyle W. and Galloway, Melanie A. and Bamford, Steven P. and others},
  title   = {Galaxy Zoo: Morphological Classifications for 120,000 Galaxies in HST Legacy Imaging},
  journal = {Monthly Notices of the Royal Astronomical Society},
  year    = {2017},
  volume  = {464},
  number  = {4},
  pages   = {4176--4203},
  doi     = {10.1093/mnras/stw2568},
}

@misc{walmsley2025gzevo,
      title={Galaxy Zoo Evo: 1 million human-annotated images of galaxies}, 
      author={Mike Walmsley and Steven Bamford and Hugh Dickinson and Tobias Géron and Alexander J. Gordon and Annette M. N. Ferguson and Lucy Fortson and Sandor Kruk and Natalie Lines and Chris J. Lintott and Karen L. Masters and Robert G. Mann and James Pearson and Hayley Roberts and Anna M. M. Scaife and Stefan Schuldt and Brooke Simmons and Rebecca Smethurst and Josh Speagle and Kyle Willett},
      year={2025},
      eprint={2512.23691},
      archivePrefix={arXiv},
      primaryClass={astro-ph.IM},
      url={https://arxiv.org/abs/2512.23691}, 
}

@inproceedings{koblischke2026galaxywiki,
  author    = {Koblischke, Nolan and Bovy, Jo},
  title     = {Extracting Published Discussions of 13,616 Galaxies with Language Models},
  booktitle = {Proceedings of the 4th Workshop on Artificial Intelligence for Scientific Publications ({WASP} 2026) at {AACL-IJCNLP} 2026},
  year      = {2026},
  url       = {https://openreview.net/forum?id=I5IVJ4QSO9}
}

\appendix
\section{Prompts}
\label{app:prompts}

We provide the exact prompts used for the \textit{Likert} and \textit{Tournament} ranking methods. For both methods, responses were returned in a predefined JSON format for consistent automated parsing.

\paragraph{\textit{Likert}.}
\begin{small}
\begin{verbatim}
You are an expert astronomer evaluating a 4x4 grid of astronomical
images. The images are indexed 1-4 in the top row, 5-8 in the second
row, 9-12 in the third row, and 13-16 in the bottom row.
Your task:
1. Select any images that are scientifically interesting. Skip blank,
   empty, or uninformative frames.
2. For each selected image, assign a Score from 1 to 5 (no explanation
   needed):
   1 = mildly interesting, 2 = somewhat interesting, 3 = interesting,
   4 = very interesting, 5 = exceptionally interesting.
Return a list of selected images. If none are interesting, return an
empty list.
\end{verbatim}
\end{small}

\paragraph{\textit{Tournament}.}
\begin{small}
\begin{verbatim}
You are an expert astronomer evaluating a 2x2 grid of four astronomical
images, labeled Image 1 (top-left), Image 2 (top-right), Image 3
(bottom-left), and Image 4 (bottom-right). Evaluate each image
independently. Do not force a winner. For each image, decide whether it
is scientifically interesting enough to keep for human inspection. It is
acceptable to keep any number of the four images (none, some, or all).
Interesting images may show unusual morphology, asymmetry, interactions,
merger-like features, arcs, rings, tails, shells, clumps, distortions,
rare-looking objects, or other features worth human inspection. Reject
artifacts, blank images, noisy frames, and obvious non-astronomical
defects. Return only whether to keep each of Image 1, Image 2, Image 3,
and Image 4. No explanation needed.
\end{verbatim}
\end{small}

\section{High-scoring candidates}
\label{app:candidates}

\begin{figure}[p]
  \centering
  \includegraphics[width=\linewidth]{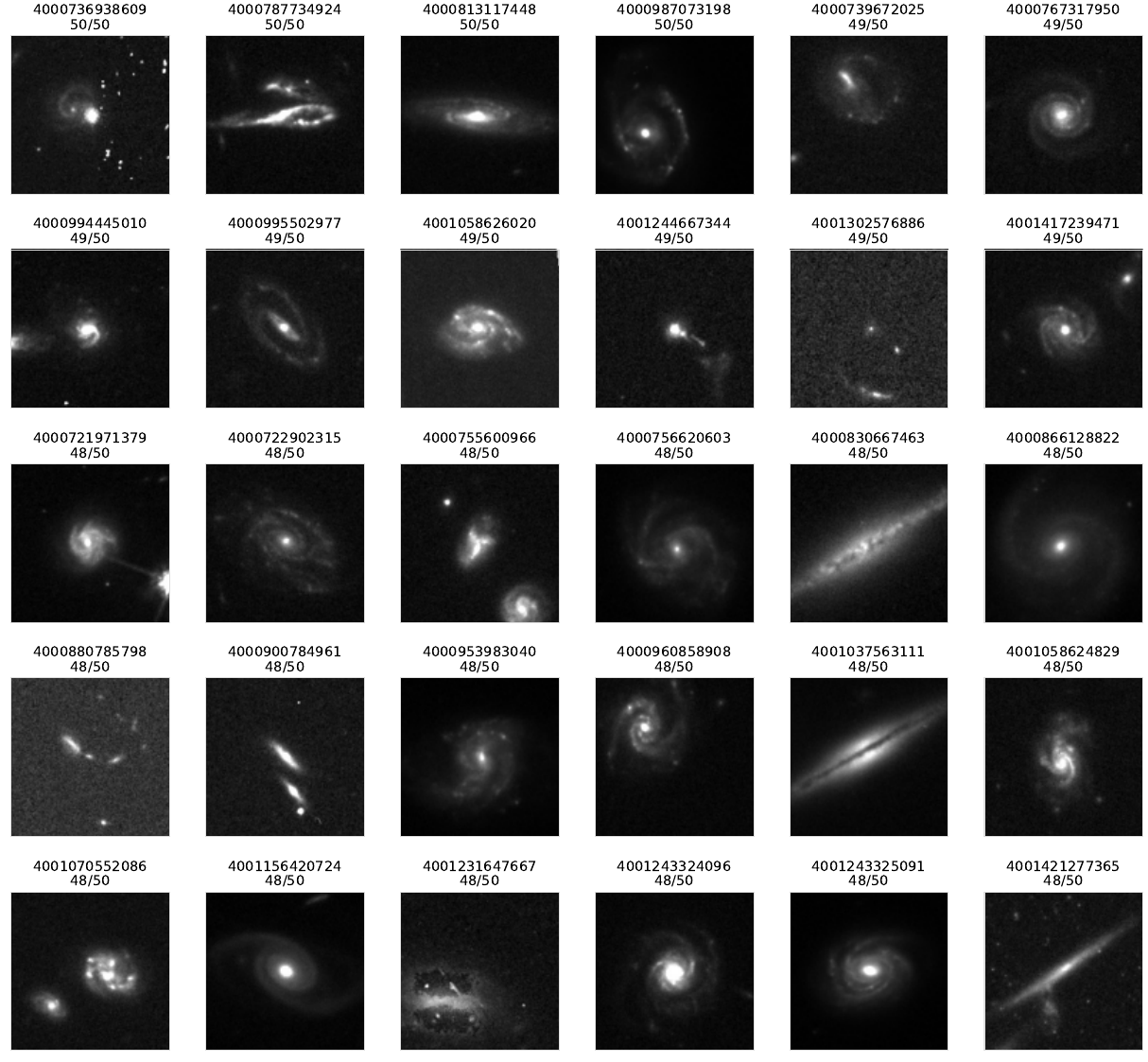}
  \caption{The 30 highest-scoring images among the 102 retained candidates with Gemini \textit{Likert} scores $\geq45$ outside AnomalyMatch, after excluding matches to existing literature catalogs within $3\arcsec$ and two sources identified as previously studied through manual literature review.}
  \label{fig:candidates}
\end{figure}

Our catalog contains the 102 high-scoring sources that remain after the catalog matching and manual literature review described above, providing a set of candidates for further astronomical investigation. Figure~\ref{fig:candidates} shows the 30 highest-scoring retained sources, ordered by decreasing score.

\section{Token usage and cost}
\label{app:cost}
Table~\ref{tab:cost} reports the token usage and estimated inference cost of each model-method combination, providing additional detail on the computational efficiency comparisons.
\begin{table}[H]
  \centering
  \caption{Token usage and cost for the four runs in Table~\ref{tab:recall}.
    Token counts are per image, averaged over each 20,000-image run. Dashes denote
    disabled reasoning for Qwen, whose output is the reported completion-token count.
    Costs are USD per 10,000 images.}
  \label{tab:cost}
  \small
  \begin{tabular*}{\textwidth}{@{\extracolsep{\fill}}llrrrr@{}}
    \toprule
    Model & Method & Input & Output & Thinking & \$/10,000 \\
    & & & & & Images \\
    \midrule
    Gemini 3.1 Flash-Lite & \textit{Likert} & 796.2 & 131.5 & 72.3 & \$5.05 \\
    Qwen3.5-397B-A17B & \textit{Likert} & 551.8 & 115.3 & -- & \$6.69 \\
    Gemini 3.1 Flash-Lite & \textit{Tournament} & 899.8 & 25.4 & 68.8 & \$3.66 \\
    Qwen3.5-397B-A17B & \textit{Tournament} & 354.9 & 18.8 & -- & \$1.80 \\
    \bottomrule
  \end{tabular*}
\end{table}

\end{document}